\documentclass[aps,prl,showpacs,floatfix,twocolumn,longbibliography]{revtex4-1}
\usepackage{mathrsfs}
\usepackage[figuresright]{rotating}
\usepackage{amsmath}
\usepackage{amssymb}
\usepackage{siunitx}
\usepackage{graphicx}
\usepackage{color}
\usepackage{dcolumn}
\usepackage{bm}
\usepackage[breaklinks=true,colorlinks=true,linkcolor=blue,urlcolor=blue,citecolor=blue]{hyperref}
\UseRawInputEncoding   

\usepackage{longtable}

\usepackage{multirow}
\usepackage{setspace}

\usepackage{soul}

\makeatletter
\newcommand{\rmnum}[1]{\romannumeral #1}
\newcommand{\Rmnum}[1]{\expandafter\@slowromancap\romannumeral #1@}
\makeatother
\usepackage{ragged2e}

\begin{document}

\title{Possible inverse magnetic melting effect in vdW-like Kondo lattice CeSn$_{0.75}$Sb$_2$}

\author{Hai Zeng$^{1}$}
\email[]{D202280149@hust.edu.cn}
\author{Yiwei Chen$^{1}$}
\author{Zhuo Wang$^{1}$}
\author{Shuo Zou$^{1}$}
\author{Kangjian Luo$^{1}$}
\author{Yang Yuan$^{1}$}
\author{Meng Zhang$^{1}$}
\author{Yongkang Luo$^{1,2}$}
\email[]{mpzslyk@gmail.com}
\address{$^1$Wuhan National High Magnetic Field Center and School of Physics, Huazhong University of Science and Technology, Wuhan 430074, China.}
\address{$^2$State Key Laboratory of Advanced Electromagnetic Technology, Huazhong University of Science and Technology, Wuhan 430074, China.}

\date{\today}

\begin{abstract}

Given the intimate connection between magnetic orders and the interplay among multiple degrees of freedom in heavy-fermion systems, controlling and understanding the associated inverse melting effect is crucial for unveiling novel condensed-matter states and their potential applications. Here, we report the growth of single crystalline quasi-two-dimensional van-der-Waals-like (vdW-like) Kondo lattice CeSn$_{0.75}$Sb$_2$, and its physical properties by a combination of transport / magnetic / thermodynamic measurements. We find that it hosts a fragile antiferromagnetic (AFM) order and a cluster glass (CG) ground state, both of which are highly sensitive to external fields. Upon cooling under low in-plane magnetic fields, the AFM phase evolves into a polarized paramagnetic phase, either directly or indirectly through the intermediate CG phase. This process constitutes a possible inverse magnetic melting effect that restores the broken translational / rotational symmetries. Our work provides a rare paradigm of inverse magnetic melting effect in vdW-like heavy-fermion materials, and enriches the physics in conventional Kondo-lattice models.\\


\end{abstract}


\maketitle

\section{\Rmnum{1}. Introduction}

Inverse melting is a rare and intriguing phenomenon in which heating paradoxically stabilizes an ordered phase, in stark contrast to conventional thermal melting \cite{Tammann1925}. This counter-intuitive phenomenon implies that the ordered phase hosts a higher entropy than the disordered state, a scenario resembling the so-called Pomeranchuk effect in liquid $^3$He that has been widely exploited for cryogenic cooling \cite{Richardson-RMP1997}. Theoretical insights from spin models elucidated the microscopic origin of this anomaly, attributing it to a significantly higher degeneracy of the interacting ordered states as compared with the non-interacting ones, which can effectively reverse the standard energy-entropy competition \cite{Schupper-PRL2004,Schupper-PRE2005}. In the realm of strongly correlated systems, it has been proposed that the competition between nearly degenerate magnetic states or intertwined orders may proliferate abundant low-energy excitations \cite{Rozen-Nature2021,Saito-Nature2021,Zhang-PRX2022,Continentino-PRB2004,Ye-PRB2024,He-SCPMA}. Such a scenario potentially endows an ordered phase with excess entropy with respect to the  disordered state, establishing the thermodynamic conditions required for inverse melting, as exemplified by twisted bilayer graphene \cite{Rozen-Nature2021,Saito-Nature2021,Zhang-PRX2022}, heavy fermion metals \cite{Continentino-PRB2004,Ye-PRB2024,He-SCPMA}, etc. In heavy-fermion systems, the competition between the Ruderman-Kittel-Kasuya-Yosida (RKKY) interaction and the Kondo effect \cite{Doniach}, as well as the interplay among spin / orbit / charge degrees of freedom \cite{Gegenwart-2008,Tsujimoto-PrV2Al20_Quadrupole,Kobayashi-YbAlB4_Mossbauer}, in principle, could create a manifold of nearly degenerate configurations as the source for extra entropy. Consequently, these systems exhibit immense potential for exploring the physics of inverse melting and may also inspire the design of functional quantum materials with electronic, magnetic and cryogenic applications.

To realize the inverse melting, it is necessary to drive the system by non-thermal tuning parameters such as magnetic field and hydrostatic pressure. A promising route for this is through the manipulation of spin dynamics in systems with high anisotropies. In this context, the intercalated Ce$Tm$Sb$_2$ ($Tm$ = transition metals) family offers an exceptional platform due to its remarkable structural tunability. Through the substitution of the transition metal $Tm$, the lattice ratio $2\tilde{c}/(\tilde{a}+\tilde{b})$ varies between 2.07 and 2.65 [seeing more details in \textbf{Supplementary Materials} (\textbf{SM}) \cite{SM}], driving the system with a more three-dimensional magnetic exchange towards a two-dimensional (2D) limit \cite{Rosa-PRB2015,Thamizhavel-PRB2003,Seo-PRX2020,Jang-PRB2024,Muro-JAC1997,Park-PRB2005}. Accompanied with this modification in structural dimensionality, the crystal electric field (CEF) ground state alternates between a configuration dominated by $|\pm5/2\rangle$ doublets (whose $4f$ spatial distribution is oblate, i.e. donut shaped) and the one characterized by $|\pm1/2\rangle$ (dumbbell shaped) \cite{Rosa-PRB2015,Willers-Ce115CEF, Cheng-CeCo2Ga8_PRM2019}. This transformation directly reshapes the spatial distribution of the $4f$ orbitals and the magnetic anisotropy, thereby regulating the sign and direction of the RKKY interaction. According to an earlier simulation based on a mean-field model including two anisotropic RKKY interactions between nearest neighbors in the presence of tetragonal CEF, the in-plane alignment of spins (with $B_2^0>0$ and the $|\pm1/2\rangle$ states dominant) is favored in ferromagnetic (FM) compounds, while $c$-axis alignment ($B_2^0<0$ and the $|\pm5/2\rangle$ state dominant) is more preferred in antiferromagnetic (AFM) compounds \cite{Rosa-PRB2015,Jang-PRB2024,Park-PRB2005}. Here $B_2^0$ is the CEF parameter of the Steven's operator $\hat{O}_2^0$ . In other words, the structural dimensionality characterized by $2\tilde{c}/(\tilde{a}+\tilde{b})$ can place the magnetic ground state in a delicate energetic balance between AFM and ferromagnetic FM orders. This competition may give rise to abundant low-energy fluctuations, which in turn establish the specific entropic landscape required for inverse melting.

\begin{figure*}[!ht]
\vspace*{5pt}
\hspace*{-0pt}
\includegraphics[width=16.0cm]{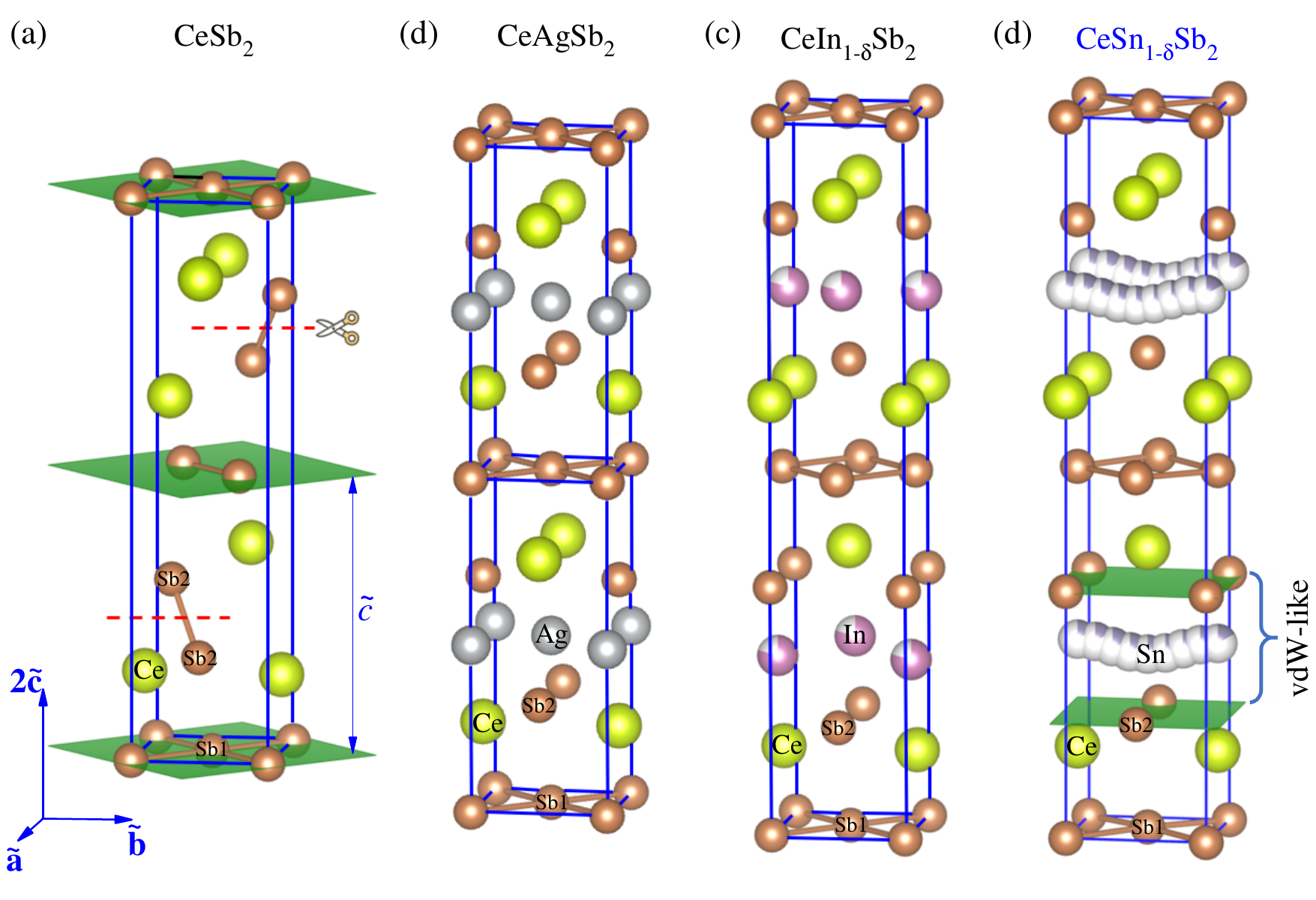}
\vspace*{-10pt}
\caption{Crystalline structures of the parent CeSb$_2$ and the intercalated Ce$Tm$Sb$_2$ series. For a better comparison, the ``unit" cell of CeAgSb$_2$ is doubled along $\tilde{\mathbf{c}}$, while those of CeSb$_2$ and CeIn$_{1-\delta}$Sb$_2$ are also modified in a proper manner, seeing Figure S1 for more details. Side views: (a) CeSb$_2$; (b) CeAgSb$_2$; (c) CeIn$_{0.8}$Sb$_2$; (d) CeSn$_{0.75}$Sb$_2$. For CeSn$_{0.75}$Sb$_2$, $\tilde{\mathbf{a}}=\mathbf{a}$, $\tilde{\mathbf{b}}=\mathbf{b}$, and $\tilde{\mathbf{c}}=\mathbf{c}/2$. $\tilde{c}$ characterizes the distance between adjacent Sb1 square nets (green squares), while $\tilde{a}$ and $\tilde{b}$ are the in-plane lattice parameters. The lattice parameters of these compounds are listed in Tab.~S1.}
\label{Fig1}
\end{figure*}

Here, we report the successful synthesis of single-crystalline CeSn$_{0.75}$Sb$_2$ and a systematic investigation into its physical properties. At zero field, CeSn$_{0.75}$Sb$_2$ exhibits a fragile AFM order below $\sim$7.5 K, followed by a transition into a cluster glass state at around 2.7 K, characterized by the coexistence of FM clusters embedded within an AFM matrix. Applying a tiny magnetic field (less than 100 Oe) along the $\mathbf{a}$-axis is sufficient to align these clusters, whereas slightly higher fields (less than 0.2 T) drive the AFM order through a metamagnetic transition (MMT) into a polarized state. A complete temperature-magnetic field phase diagram of CeSn$_{0.75}$Sb$_2$ is mapped out, and a possible inverse melting of the magnetic order is observed by transport, magnetic and therodynamic measurements. These results suggest a potential mechanism for applications in quantum switches, memory devices or solid-state cryogen.

\section{\Rmnum{2}. Experimental details}

\begin{figure*}[!ht]
\vspace{5pt}
\hspace{-0pt}
\includegraphics[width=16.0cm]{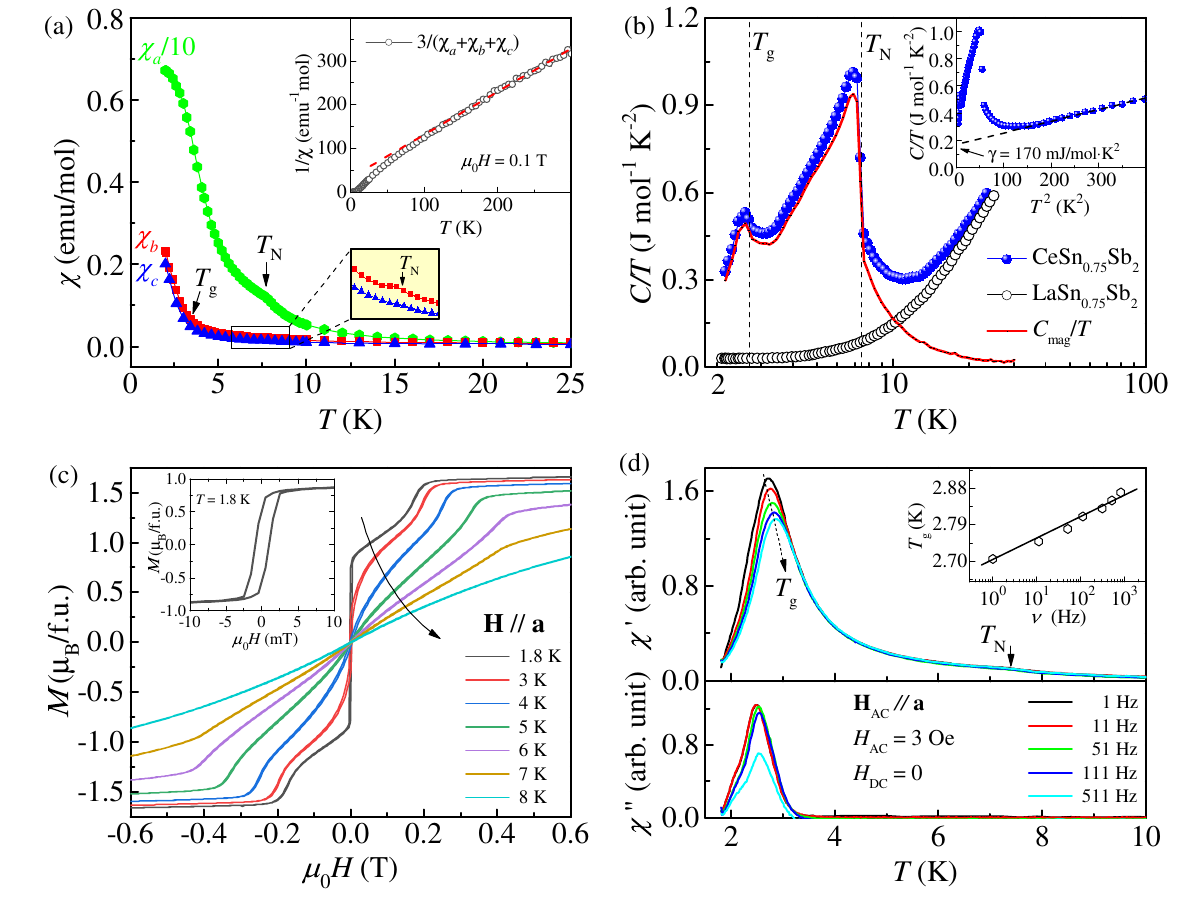}
\vspace*{-10pt}
\caption{(a) Temperature dependence of magnetic susceptibility $\chi$ for different field orientations measured with $\mu_0H=$ 0.1 T. The inset shows the temperature dependence of $1/\chi_\text{avg}$, where $\chi_\text{avg}=(\chi_a+\chi_b+\chi_c)/3$. The red dashed line is a fit to the Curie-Weiss law. (b) Specific heat $C/T$ curves of CeSn$_{0.75}$Sb$_2$ and its nonmagnetic analog LaSn$_{0.75}$Sb$_2$. The red solid line represents the magnetic contribution $C_\text{mag}/T$. The inset exhibits the $C/T$ vs. $T^2$ plot whose intercept of the linear extrapolation yields the Sommerfeld coefficient $\gamma=170$ mJ/mol$\cdot$K$^2$. (c) Field dependence of the magnetization $M(H)$ at various temperatures for $\mathbf{H}\parallel\mathbf{a}$. The inset displays the magnetic hysteresis loop about zero field at 1.8 K. (d) Real ($\chi^{\prime}$) and imaginary ($\chi^{\prime \prime}$) parts of AC magnetic susceptibility vs temperature at different frequencies. The applied AC excitation field, $H_\text{AC}$, was along $\mathbf{a}$. The inset shows $T_\text{g}$ as a function of frequency $\nu$.  }
\label{Fig2}
\end{figure*}

Single crystals of CeSn$_{0.75}$Sb$_2$ were synthesized by the self-flux method. Ce flakes (99.9\%), Sn granules (99.999\%), and Sb grains (99.99\%) were mixed in a molar ratio of 1 : 10 : 2.5, placed in an alumina crucible, and then sealed in a high-vacuum quartz tube. The mixture was heated to 950 $^\circ$C in 24 h, held for 12 h, and then slowly cooled to 500 $^\circ$C at a rate of 2 $^\circ$C/h. The Sn flux was subsequently removed by centrifugation at this temperature. Flake-shaped CeSn$_{0.75}$Sb$_2$ single crystals were obtained, with a typical size of 3$\times$0.5$\times$0.02 mm$^3$. The sample is stable in air. LaSn$_{0.75}$Sb$_2$ single crystals, the non-$4f$ reference, were synthesized in a similar method. The energy-dispersive X-ray spectroscopy (EDS) analysis confirmed the stoichiometric ratio of Ce, Sn, and Sb as 1.03 (1) : 0.75 (1) : 2.02 (1). The phase purity and high crystalline quality were verified by X-ray diffraction (XRD) using an XtaLAB mini \Rmnum{2} single crystal diffractometer, where only the [0 0 $2l$] ($l=1,2...$) peaks are visible (Fig.~S2 \cite{SM}).

Magnetic measurements, including both direct-current (DC) and alternating-current (AC) magnetization, were performed using a vibrating sample magnetometer (SQUID-VSM, Quantum Design). Zero-field specific heat measurements were performed using a commercial Physical Property Measurement System (PPMS-9, Quantum Design). Thermodynamic properties under external magnetic field were characterized by AC calorimetry ($C_\text{ac}$), which was measured by using a differential Chromel-Au$_{99.93\%}$Fe$_{0.07\%}$ thermocouple to detect the heat-temperature response. Electrical resistivity measurements were carried out in a standard four-probe method with applied magnetic field along $\mathbf{a}$ axis. Both AC calorimetry and electrical transport experiments were conducted in an IntegraAC Mk \Rmnum{2} recondensing cryostat equipped with a 16 T superconducting magnet (Oxford Instruments).

\section{\Rmnum{3}. Results and Discussion}

The material we investigated here, CeSn$_{0.75}$Sb$_2$, belongs to the large family of Ce$Tm$Sb$_2$ ($Tm$ = intercalated metals) that can be viewed as $Tm$-intercalation in CeSb$_2$, as expressed in Figure \ref{Fig1}. The parent compound CeSb$_2$ crystallizes in an orthorhombic structure with alternating layers of Sb1 square nets and Ce-Sb2 slabs stacked along the $\tilde{\mathbf{c}}$-axis \cite{Wang-IC1967}. The $Tm$ intercalation takes place by breaking the chemical bonds between Sb2 atoms [cf. Fig.~\ref{Fig1}(a)]. Due to the new bondings between $Tm$-Sb2, the position of Ce atoms are rearranged correspondingly, and eventually this leads to various types of structures with a common chemical formula, Ce$Tm$Sb$_2$ (e.g., $Tm$ = Ag, In, Sn, etc) \cite{Ferguson-IC1999, Araki-PRB2003}. Among them, CeSn$_{0.75}$Sb$_2$ is special in that Sn atoms distribute over three crystallographically distinct sites, each partially occupied by about 20\% \cite{Ferguson-IC1996}, and this results in the worm-like Sn chains along $\tilde{\mathbf{b}}$ [Fig.~\ref{Fig1}(d)]. More strinkingly, previous M\"{o}ssbauer spectroscopy study on polycrystalline samples revealed that the intercalated Sn atoms are essentially zero-valent with little charge transfer \cite{Deakin-JSSC2002}. This renders the weak interlayer coupling, and indeed, the obtained CeSn$_{0.75}$Sb$_2$ single crystals appear van-der-Waals-like (vdW-like) and can be easily exfoliated mechanically. A natural consequence of such intercalation is that the $\mathbf{c}$ axis is elongated and the systems becomes more anisotropic in nature, as characterized by the enlarged ratio $2\tilde{c}/(\tilde{a}+\tilde{b})\approx2.65$ as shown in Tab.~S1 (The definitions of $\tilde{a}$, $\tilde{b}$ and $\tilde{c}$ are explained in Fig.~S1 \cite{SM}). Such a ratio significantly surpasses the critical AFM¨CFM boundary ($\approx $2.37), placing it deeply inside the two-dimensional regime where the $|\pm1/2\rangle$ doublet is expected to be dominant in the CEF ground state \cite{Rosa-PRB2015}. This extreme orbital dimensionality not only attenuates the interlayer antiferromagnetic coupling but also renders the magnetic ground state exceptionally susceptible to thermal and magnetic perturbations, potentially meeting the conditions for inverse melting.

Figure \ref{Fig2}(a) presents the temperature dependent magnetic susceptibility $\chi (T)$ of CeSn$_{0.75}$Sb$_2$ single crystal under external field 0.1 T along different crystallographic axes. In order to compare with the earlier polycrystalline sample \cite{Deakin-JSSC2002}, the powder-averaged magnetic susceptibility [$\chi_\text{avg}\equiv(\chi_a+\chi_b+\chi_c)/3$] is displayed in the inset. Curie-Weiss's law is well obeyed above 150 K, consistent with the literature. At low temperature, $\chi_a$, $\chi_b$ and $\chi_c$ behave rather differently. To our surprise, a huge in-plane magnetic anisotropy is observed: at 1.8 K, $\chi_{a}$ is about 30 times larger than $\chi_b$, whereas the values of $\chi_{b}$ and $\chi_{c}$ are comparable. Isothermal field dependent magnetization [$M(H)$] further confirms $\mathbf{a}$ as the easy axis, seeing Fig.~S3(b) \cite{SM}. Such a giant in-plane anisotropy indicates an Ising-type magnetic character, which has rarely been reported in quasi-2D materials. Note that the lattice parameter $b$ is only $6.6\%$ larger than $a$ (Tab.~S1 \cite{SM}). We speculate that the Sn chains along $\mathbf{b}$ probably play a key role for this in-plane anisotropic behavior. Since the magnetic anisotropy of Ce-based compounds is usually governed by the CEF effect, further investigations into the CEF-split $4f$ orbitals will be essential to clarify the origin of this unusual anisotropy. A kink is clearly seen in $\chi_a(T)$ around 7.5 K, followed by a rapid increase for temperatures below 3 K, suggesting cascaded magnetic transitions in this compound, and this is further supported by the specific heat results shown in Fig.~\ref{Fig2}(b). The critical temperatures of the two transitions are defined as $T_\text{N}=7.5$ K and $T_\text{g}=2.7$ K, respectively. The two transitions are also discernible in $\chi_b(T)$, whereas $T_\text{N}$ is hardly resolvable in $\chi_c(T)$. It is worthwhile to mention that an earlier study on polycrtstalline CeSn$_{0.7}$Sb$_2$ unveiled the only transition around $T_\text{g}$ \cite{Deakin-JSSC2002}. 
All these suggest that the magnetic behavior is governed primarily by strong in-plane exchange interactions, with interlayer coupling remaining comparatively weak.

\begin{figure}[!ht]
\vspace*{5pt}
\hspace*{-0pt}
\includegraphics[width=8.5cm]{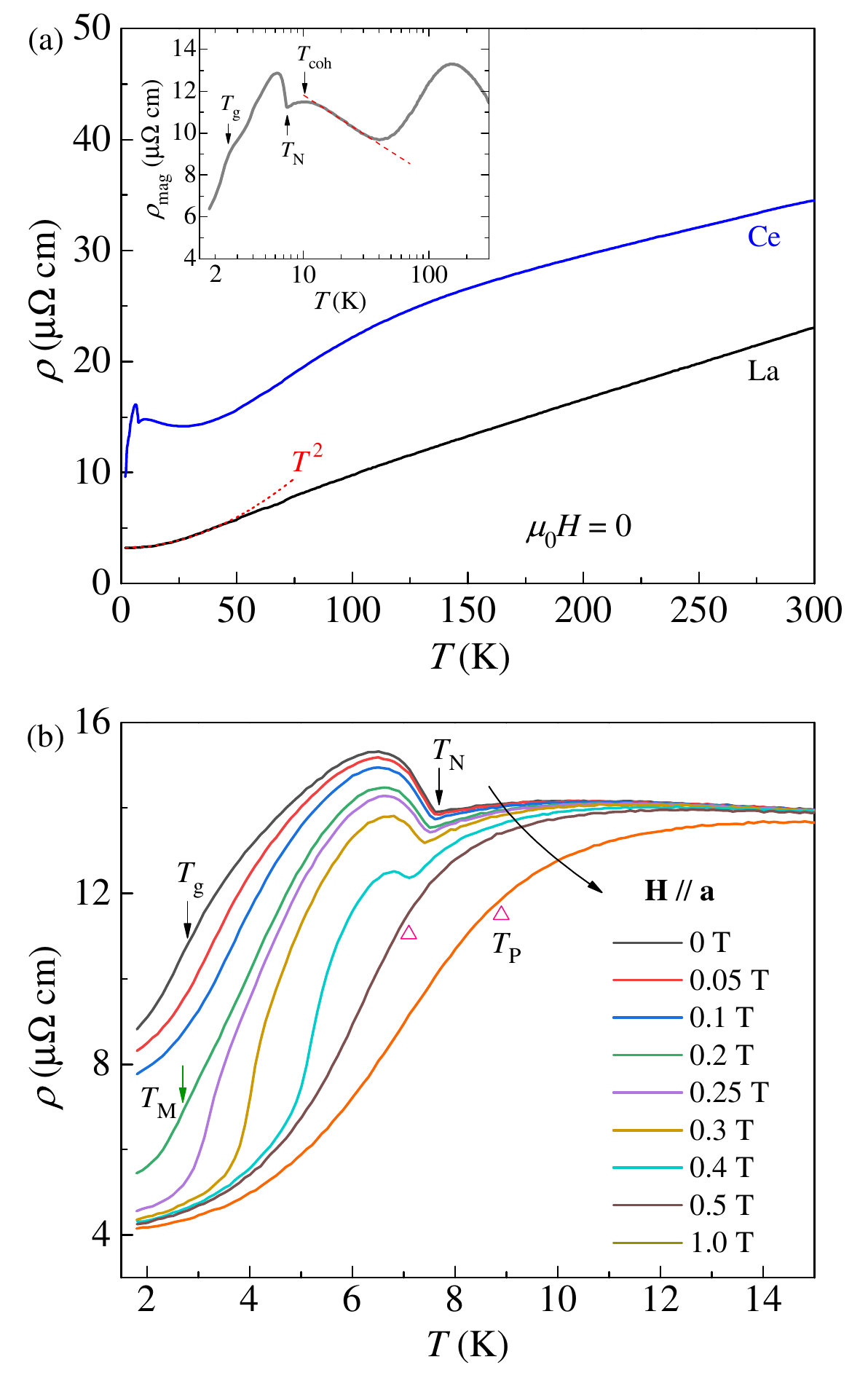}
\vspace*{-20pt}
\caption{Longitudinal resistivity of CeSn$_{0.75}$Sb$_2$. (a) $\rho(T)$ profile of CeSn$_{0.75}$Sb$_2$ as compared with that of LaSn$_{0.75}$Sb$_2$. The inset displays the contribution of the 4$f$-electrons to the resistivity, revealing the coherent Kondo scale $T_\text{coh}$, and the two magnetic transitions at $T_\text{N}$ and $T_\text{g}$. (b) The temperature dependence of the resistivity under various magnetic fields with $\mathbf{H}\parallel\mathbf{a}$ and $\mathbf{I}\parallel\mathbf{a}$. The open triangles signify the characteristic temperature $T_\text{P}$.  }
\label{Fig3}
\end{figure}

\begin{figure*}[!ht]
\vspace*{5pt}
\includegraphics[width=16.5cm]{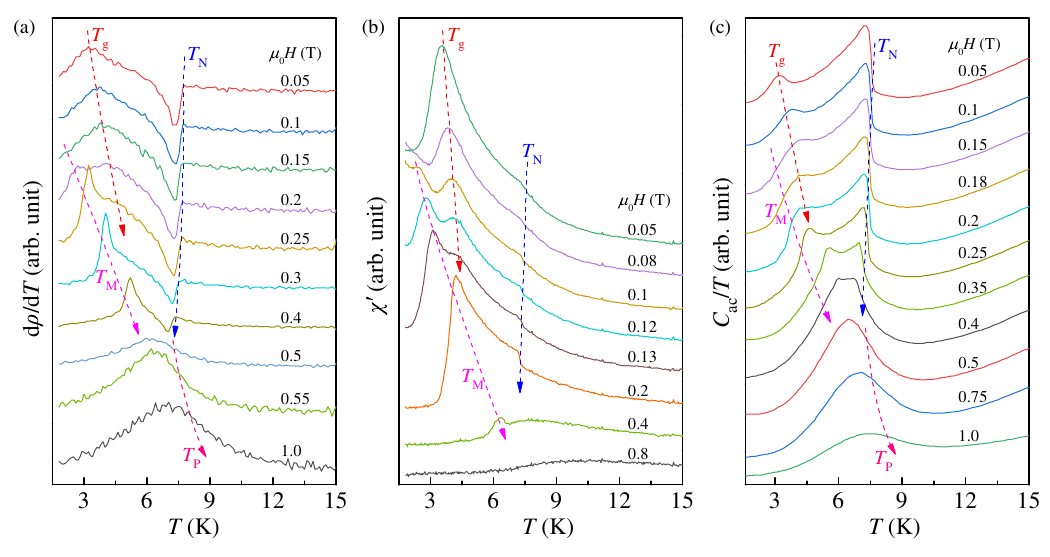}
\vspace*{-10pt}
\caption{Temperature dependence of $d\rho/dT$ (a), $\chi^\prime$ (b), and $C_\text{ac}/T$ (c). For all these measurements, both $H_\text{DC}$ and $H_\text{AC}$ were applied along $\mathbf{a}$. The AC susceptibility was measured with an excitation amplitude of 3.8 Oe at a frequency of 511 Hz, while the AC specific heat was recorded with a modulation frequency of 51 Hz. The curves are shifted vertically for clarity. $T_\text{N}$ (blue arrows), $T_\text{g}$ (red arrows), and $T_\text{M}$ (pink arrows) denote the AFM, CG, and metamagnetic transition temperatures, respectively. }
\label{Fig4}
\end{figure*}

To further investigate the features of the two magnetic transitions, we carried out additional magnetization measurements at low temperature. The rapid increase of $\chi(T)$ below $T_\text{g}$ is reminiscent of a FM-like correlation, and this seems to be supported by the bifurcation between zero-field-cooling (ZFC) and field-cooling (FC) protocols under small fields less than 80 Oe [cf Fig.~S3(a) \cite{SM}], as well as the small hysteresis loop in isothermal magnetization $M(H)$ curves shown in Fig.~\ref{Fig2}(c). Such an irreversible behavior, however, may be also a typical signature of systems with frozen magnetic disorder, e.g. spin glasses / cluster glass (CG) \cite{Binder-1986,Wu-PRB2003,Luo-Li2RhO3_PRB2013}, and superparamagnetic assemblies \cite{Bedanta-JPD2009}. To further clarify this behavior, we measured the temperature dependence of AC susceptibility at multiple frequencies ($\nu$) with fixed $\mathbf{a}$-axis excitation field $H_\text{AC}=3$ Oe [Fig.~\ref{Fig2}(d)]. The real part ($\chi^{\prime}$) reveals two distinct features: a sharp peak near 2.7 K and a tiny hump around 7.5 K. The imaginary part ($\chi^{\prime\prime}$), however, only uncovers the dissipation peak around 2.7 K. The absence of dissipation at $T_\text{N}$ is consistent with the second-order nature of the AFM transition. More crucially, the peak near 2.7 K exhibits a pronounced frequency dependence [inset of Fig.~\ref{Fig2}(d)], a hallmark of spin-glass or cluster-glass dynamics. The corresponding Mydosh parameter ($\Phi = \frac{\Delta T_\text{g}}{T_\text{g} \Delta \log \nu}$) is determined to be $\sim0.023$. This value is in better agreement with typical CG systems ($\Phi\sim0.03$) than with canonical spin-glass systems ($\Phi\sim0.005$) \cite{Mydosh-RPP2015, Xiao-JAC2020, Nath-PRB2018}. Therefore, the observed irreversibility in the susceptibility curves is more likely a consequence of CG state, wherein randomly frozen FM clusters are embedded inside a disordered AFM matrix, and different cooling paths can lead to distinct cluster orientations. This configuration probably originates from the local crystalline disorder induced by the partial occupancy of Sn at three independent sites. Unlike canonical metallic spin glasses where random placement of magnetic moments generates sign-oscillating RKKY interactions, the chemical disorder alters the local ligand environment around Ce ions, which theoretically could lead to a spatial modulation of the hybridization between Ce-4f and conduction electrons. Such hybridization fluctuations may interact with the randomness in RKKY exchange, promoting the formation of a cluster-glass state. A similar mechanism has been established in other heavy-fermion systems, such as CePd$_{1-x}$Rh$_x$ \cite{Westerkamp-CePd_Rh_Cluster_glass} and URh$_2$Ge$_2$ \cite{Menovsky-PRL1997}, where mixed occupation of ligand sites produces a broad distribution of local Kondo temperatures.

The magnetic transitions in CeSn$_{0.75}$Sb$_2$ are also seen in electrical transport properties, as shown in Fig.~\ref{Fig3}. For comparison, the non-$4f$ counterpart LaSn$_{0.75}$Sb$_2$ is also shown in the same frame. LaSn$_{0.75}$Sb$_2$ behaves as a simple metal, viz $\rho$ decreases almost linearly with decreasing $T$ for above 100 K, and then crossover into a Fermi liquid behavior [$\rho \propto T^2$, cf the dashed line in Fig.~\ref{Fig3}(a)] at low temperature. CeSn$_{0.75}$Sb$_2$ also shows a metallic profile, but additional anomalies are observable. First, a broad hump is seen around 120 K, which can be more clearly identified by subtracting the resistivity of LaSn$_{0.75}$Sb$_2$ [cf $\rho_\text{mag}$ shown in the inset to Fig.~\ref{Fig3}(a)]. Such kind of broad peak in $\rho_\text{mag}(T)$ is commonly seen in Ce-based compounds and should be ascribed to the scatterings between the thermally populated CEF levels \cite{Bauer-CEF, LuoY-CeNi2As2}. Further decreasing temperature to below 50 K, as incoherent Kondo scattering sets in, the resistivity turns up and increases logarithmically until a second broad peak centering about 10 K is observed. This implies the establishment of Kondo coherence and yields an important temperature scale for Kondo lattices, $T_\text{coh}\approx10$ K. Around $T_\text{N}$, resistivity increases sharply, passing through a peak and then decreases rapidly. Such kind of $\rho(T)$ profile near AFM transition is generally attributed to a competition between suppressed spin scattering rates and reduced carrier concentration as a consequence of gap opening on the Fermi surface upon AFM ordering. In fact, the reduction of carrier density is also supported by the Hall effect as shown in Fig.~S4. One possibility may be that this magnetic transition is also accompanied with a kind of density-wave order, which deserves further investigation. 
A distinct kink is observed near 2.7 K ($\sim T_\text{g}$), corresponding to the low-temperature magnetic cluster transition. As shown in Fig. 3(b), an application of magnetic field initially suppresses $T_\text{N}$, and the sharp resistivity change near $T_\text{N}$ becomes progressively smeared out. For field larger than 0.5 T, no anomaly in resistivity is observed at this region, but instead, $\rho(T)$ decreases smoothly and rapidly. This indicates that the magnetic moments tend to be polarized by external field prior to undergoing the AFM transition, which can significantly reduce the spin scatterings. Similar behavior was also reported in CeAlGe \cite{He-SCPMA}. A new characteristic temperature $T_\text{P}$ is thus defined to describe the polarization of Ce moments under field. Interestingly, when field is larger than 0.2 T, an additional magnetic transition emerges at $T_\text{M}$$\sim$2.65 K. Both $T_\text{g}$ and $T_\text{M}$ tend to increase with magnetic field, which can be better seen in $d\rho/dT$ [Fig.~\ref{Fig4}(a)] and will be discussed further in the following.

Turning now back to the isothermal magnetization results of CeSn$_{0.75}$Sb$_2$. Another prominent feature of the $M(H)$ curves is that they show slope changes along different crystallographic directions at 1.8 K, seeing Fig.~S3(b). In particular, for field along $\mathbf{a}$, a field-induced metamagnetic transition (MMT) is clearly seen near 0.17 T [Fig.~\ref{Fig2}(c)]. Above this critical field, the magnetization levels off. The saturated moment is 1.68 $\mu_\text{B}$/f.u.  Unexpectedly, as temperature increases, the critical field for this MMT moves to higher values; meanwhile, the signature of MMT gradually dies away and becomes indistinguishable for temperatures above $T_\text{N}$. First, the observation of this field-induced MMT affirms that the ground state of CeSn$_{0.75}$Sb$_2$ at $T_\text{g}<T<T_\text{N}$ is a kind of antiferromagnetic ordering, albeit that the exact magnetic structure awaits to be determined by other experiments such as neutron scattering. Second, the upward movement of the critical field with increasing temperature itself is rather unusual, which directly means that the magnetization (or reorientation) of magnetic moments becomes even harder at elevated temperatures. Such a behavior is counter-intuitive and reminds us of the possibility that an inverse melting of the magnetic ordering may take place in this system.

In order to gain deeper insights into the inverse melting phenomenon, systematic resistivity ($d\rho/dT$), the real part of the AC magnetic susceptibility ($\chi^\prime$), and AC calorimetry ($C_\text{ac}/T$) measurements (note \cite{AC_calorimetry}) were conducted on CeSn$_{0.75}$Sb$_2$ under magnetic field along $\mathbf{a}$ axis, and the results are displayed in Fig.~\ref{Fig4}. $T_\text{N}$ shifts weakly to lower temperatures with increasing magnetic field, consistent with AFM order. The evolution of $T_\text{g}$ with field is opposite to that of $T_\text{N}$. When field is larger than $\sim0.1$ T, both $d\rho/dT$ and $\chi^\prime$ show a distinct peak near $T_\text{M}$ associated with the metamagnetic transition, whereas $C_\text{ac}/T$ displays only a broad shoulder. $T_\text{M}$ increases with field faster than $T_\text{g}$, and finally the two merge when field exceeds $\sim$0.2 T, reforming a sharp peak. As field is further increased, the Ce moments tend to be polarized, therefore $T_\text{N}$, $T_\text{g}$ and $T_\text{M}$ are no longer distinguishable. In general, the three measurements exhibit consistency in characterizing the magnetic transitions. Under identical magnetic fields (e.g., 0.1 T, 0.2 T, and 0.4 T), the $T_\text{N}$ determined by different techniques are in good agreement, whereas $T_\text{g}$ and $T_\text{M}$ display measurable variations. This is probably related to the glassy dynamics in the cluster-glass state, whose relaxation time spectrum leads to a frequency dependence of the freezing temperatures.

\begin{figure}[!ht]
\vspace*{5pt}
\hspace*{-10pt}
\includegraphics[width=9cm]{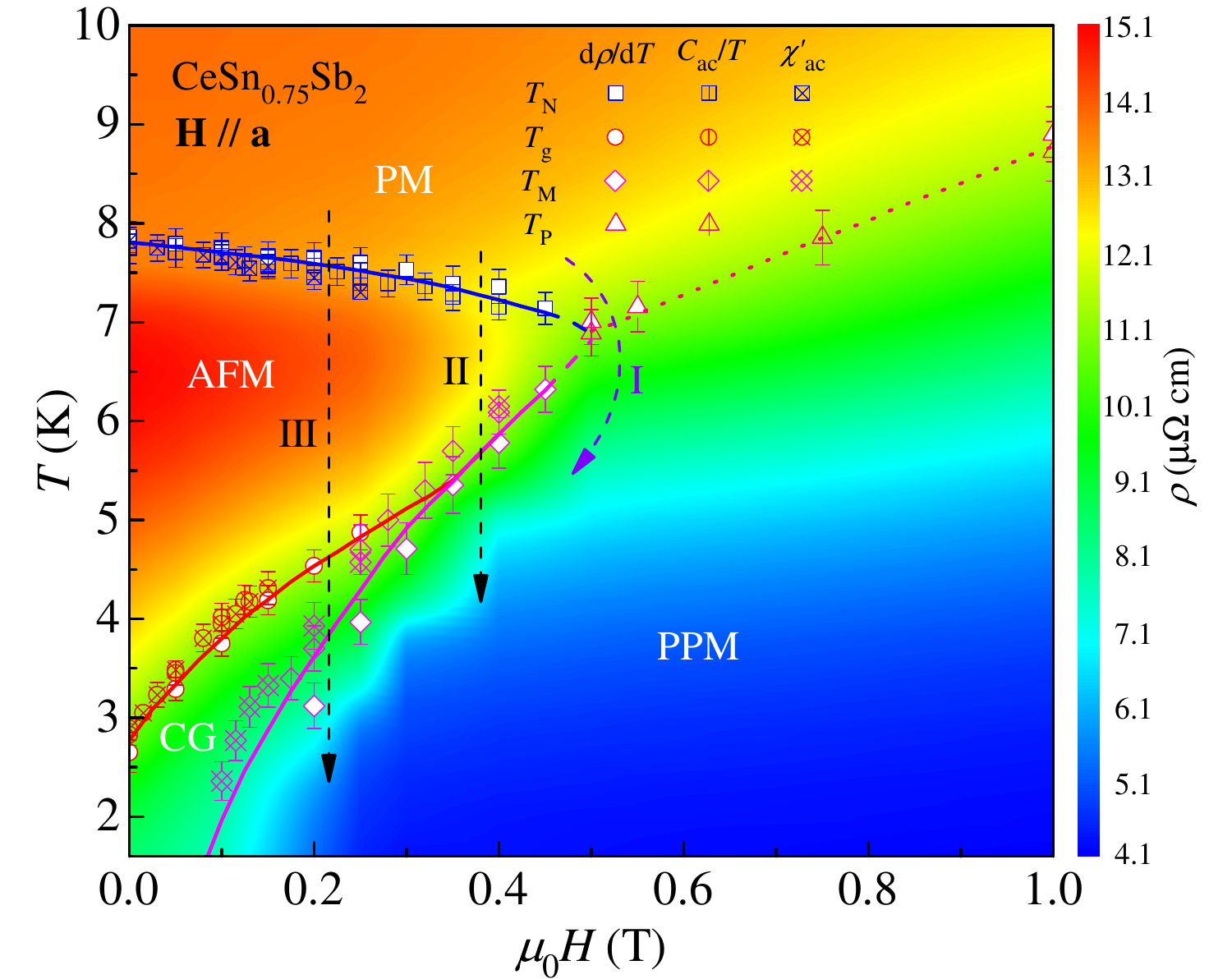}
\vspace*{-15pt}
\caption{False-color phase diagram of CeSn$_{0.75}$Sb$_2$ constructed by a contour plot of $\rho(\mu_0H, T)$, with the field applied along the $\mathbf{a}$-axis. The abbreviations are: AFM (antiferromagnetic), PM (paramagnetic), PPM (polarized paramagnetic), CG (cluster glass). The dashed arrows \Rmnum{1}-\Rmnum{3} represent different thermodynamic cooling paths from PM to PPM under variable magnetic fields. The yellow dashed line denotes the crossover from PM to field-induced PPM state. Note that cooling destabilizes the ordered state even without field or passing through the CG phase.}
\label{Fig5}
\end{figure}

Based on the results described above, a temperature-field phase diagram of CeSn$_{0.75}$Sb$_2$ for $\mathbf{H}\parallel\mathbf{a}$ is constructed in Fig.~\ref{Fig5}, where the false color indicates the magnitude of $\rho(\mu_0H, T)$. Upon cooling in zero field, the system successively enters the AFM state below $T_\text{N}$ and the CG phase below $T_\text{g}$. Applying an external magnetic field polarizes the magnetic phases (AFM and CG) into a polarized paramagnetic (PPM) state, with the transition boundary determined by $T_\text{M}$. One finds that the lines of $T_\text{N}$ and $T_\text{M}$ meet at around (0.5 T, 6.9 K). This diagram reveals three distinct cooling pathways depending on the strength of the applied magnetic field: (\rmnum{1}) For fields above $\sim 0.5$ T, the system exhibits a smooth crossover from the PM state to the PPM state, involving no distinct phase transition (Path I). Since the applied magnetic field explicitly breaks time-reversal symmetry, no additional \textit{spontaneous} symmetry breaking occurs between PM and PPM, and a phase transition is avoided. (\rmnum{2}) When cooled in a moderate field, the system  undergoes a PM-to-AFM transition followed by the formation of PPM state (Path II). Since the broken translational and rotational symmetries in the high-$T$ AFM phase are restored in the low-$T$ PPM phase, this process is identified as inverse melting \cite{Schupper-PRL2004,Schupper-PRE2005,Ye-PRB2024}. (\rmnum{3}) In the low-field regime (Path III), inverse melting also takes place upon cooling, passing through an intermediate CG phase. It is crucial to emphasize that the observed inverse magnetic melting is not a trivial consequence of field-induced polarization. If the AFM state is solely suppressed by Zeeman effect, one would anticipate a monotonic suppression of the ordered phase. However, the observed reentrant restoration of the higher-symmetry phase upon field-cooling is inconsistent with this scenario, indicating a mechanism beyond energy-driven.

Inverse melting effect has been observed in a variety of condensed-matter systems, and is generally explained as a consequence of energy-entropy competition. Besides the classic Pomeranchuk effect in liquid $^3$He, a recent example is twisted bilayer graphene \cite{Rozen-Nature2021,Saito-Nature2021}, in which upon heating, the electronic state ``freezes" from a metal phase (wherein electrons can move freely) to a near-insulating phase (in which the electrons are localized and ``ordered"). The latter is favored at elevated temperatures because of the more entropy gain by cooperating the isospin degrees of freedom \cite{Lian-Nature2021}. In spin systems, magnetic orders usually get further stabilized approaching zero temperature, due to the quench of thermal fluctuations. In the presence of magnetic field, as the Zeeman energy competes with the magnetic exchange, an AFM order is often suppressed and terminated by either a quantum critical point or a 1st-order quantum phase transition \cite{Grigera2001, Gegenwart2002, LuoY-CeNi2As2Pre, JiaoL-CeRhIn5B}. In contrast, the occurrence of inverse melting is much rarer. While the exact reason for the inverse magnetic melting effect in CeSn$_{0.75}$Sb$_2$ is to be understood, one possibility is that the emergence of the CG state out of the AFM order is associated with an enhanced degeneracy of magnetic configurations, which may ease the entropy-driven inverse melting behavior. Altogether, the AFM order manifests itself as an entropy-driven metastable phase that is confined to a finite temperature window and collapses upon cooling. Additional evidence supporting the entropy-driven mechanism can be provided by the magneto-calorimetric measurements shown in Fig.~S6 in the \textbf{SM}. These findings highlight the potential of utilizing entropy as a tuning parameter.

Finally, it is necessary to revisit the specific heat results in Fig.~\ref{Fig2}(b). A $C/T$ vs. $T^2$ plot yields the Sommerfeld coefficient $\gamma = 170$ mJ/mol$\cdot$K$^2$. By integrating $C_\text{mag}/T$ over $T$, we derive the magnetic entropy as a function of temperature, as shown in Fig.~S5 \cite{SM}, where $C_\text{mag}$ is the $4f$ magnetic contribution to specific heat. The Kondo scale ($T_\text{K}$) can thus be estimated from $S_\text{mag}(T_\text{K}/2)=0.4R\ln{2}$ \cite{Gegenwart-2008}, which results in $T_\text{K}\approx10$ K.
All these manifest that CeSn$_{0.75}$Sb$_2$ is a moderate heavy-fermion metal with relatively weak Kondo effect, sitting in the magnetically-ordered regime on the Doniach's phase diagram \cite{Doniach}. It is interesting to point out that inverse magnetic melting effect has so far be reported in a handful of Ce-based heavy-fermion materials including Ce$_3$TiSb$_5$ \cite{Ce3TiSb5-IMM} and CeAlGe \cite{He-SCPMA}. In both Ce$_3$TiSb$_5$ and CeAlGe, the AFM transitions are initially enhanced by application of pressure \cite{Shinozaki-Ce3TiSb5Pressure,He-SCPMA}, and so is the case in CeSn$_{0.75}$Sb$_2$. (The pressure effect of CeSn$_{0.75}$Sb$_2$ will be published separately \cite{Zeng-CeSnSb2_Pressure}). These commonalities tend to suggest that inverse magnetic melting effect in heavy-fermion systems are more favored in those with weak Kondo coupling. The underlying mechanism for this calls for more investigations, both experimentally and theoretically.

\section{\Rmnum{4}. Conclusions}

In summary, the vdW-like Kondo lattice CeSn$_{0.75}$Sb$_2$ exhibits a variety of magnetic phases that are highly sensitive to magnetic fields, alongside a rare and highly tunable inverse melting effect. At zero field, the system hosts a fragile AFM order and a CG ground state, which are readily suppressed by an applied in-plane magnetic field. Notably, multi-pathway inverse melting occurs upon cooling under specific magnetic fields. The microscopic origin of the inverse magnetic melting behavior remains elusive but likely arises from a thermodynamic preference for a high-entropy ordered phase. These findings not only extend the material basis of quasi-2D heavy-fermion compounds beyond CeTe$_3$ \cite{Zeng-CeTe3_Newton2026} and the recently discovered CeSiI \cite{Posey-CeSiI_Nature2024}, but also offer a fresh thermodynamic perspective for understanding the complex phases in correlated electronic systems.

\section{Acknowledgments}

The authors thank Zhentao Wang, Xiaoxiao Zhang and Yifeng Yang for helpful discussions. This work is supported by the National Key R\&D Program of China (2022YFA1602602 and 2023YFA1609600), National Natural Science Foundation of China (U23A20580 and 52588101), and Beijing National Laboratory for Condensed Matter Physics (2024BNLCMPKF004).


\begin{thebibliography}{10}
\bibitem{AC_calorimetry}
Since the CeSn$_{0.75}$Sb$_2$ single crystals are very thin, typically $\sim~20
  \mu\text{m}$, and it is very difficult to measure specific heat with in-plane
  magnetic field in PPMS. Therefore, we performed the AC calorimetry
  measurement instead.
\end{thebibliography}

\providecommand{\newblock}{}


\newpage

\renewcommand{\thefigure}{S\arabic{figure}}
\renewcommand{\thetable}{S\arabic{table}}
\renewcommand{\theequation}{S\arabic{equation}}
\onecolumngrid

\newpage

\begin{center}
{\bf \large
Supplementary Materials:\\
Possible inverse magnetic melting effect in vdW-like Kondo lattice CeSn$_{0.75}$Sb$_2$
}
\end{center}

\setcounter{table}{0}
\setcounter{figure}{0}
\setcounter{equation}{0}
\setcounter{page}{1}

\small
\begin{center}
Hai Zeng$^{1*}$\email{D202280149@hust.edu.cn}, Yiwei Chen$^{1}$, Zhuo Wang$^{1}$, Shuo Zou$^{1}$, Kangjian Luo$^{1}$, Yang Yuan$^{1}$, Meng Zhang$^{1}$, and Yongkang Luo$^{1,2\dag}$\email{mpzslyk@gmail.com}\\
$^1${\it Wuhan National High Magnetic Field Center and School of Physics, Huazhong University of Science and Technology, Wuhan 430074, China;}\\
$^2${\it State Key Laboratory of Advanced Electromagnetic Technology, Huazhong University of Science and Technology, Wuhan 430074, China.}\\

\date{\today}
\end{center}
\normalsize
\vspace*{15pt}

In this \textbf{Supplementary Materials} (\textbf{SM}), we provide additional results that further support the discussion and conclusion in the main text, including structural analysis, sample characterization, magnetization, Hall effect, and magnetic entropy. \\

\newpage

\section{SM \Rmnum{1}. S\lowercase{tructural information about \uppercase{C}e$\uppercase{T}m$\uppercase{S}b$_2$}}

\begin{figure*}[!ht]
\vspace*{-0pt}
\includegraphics[width=15cm]{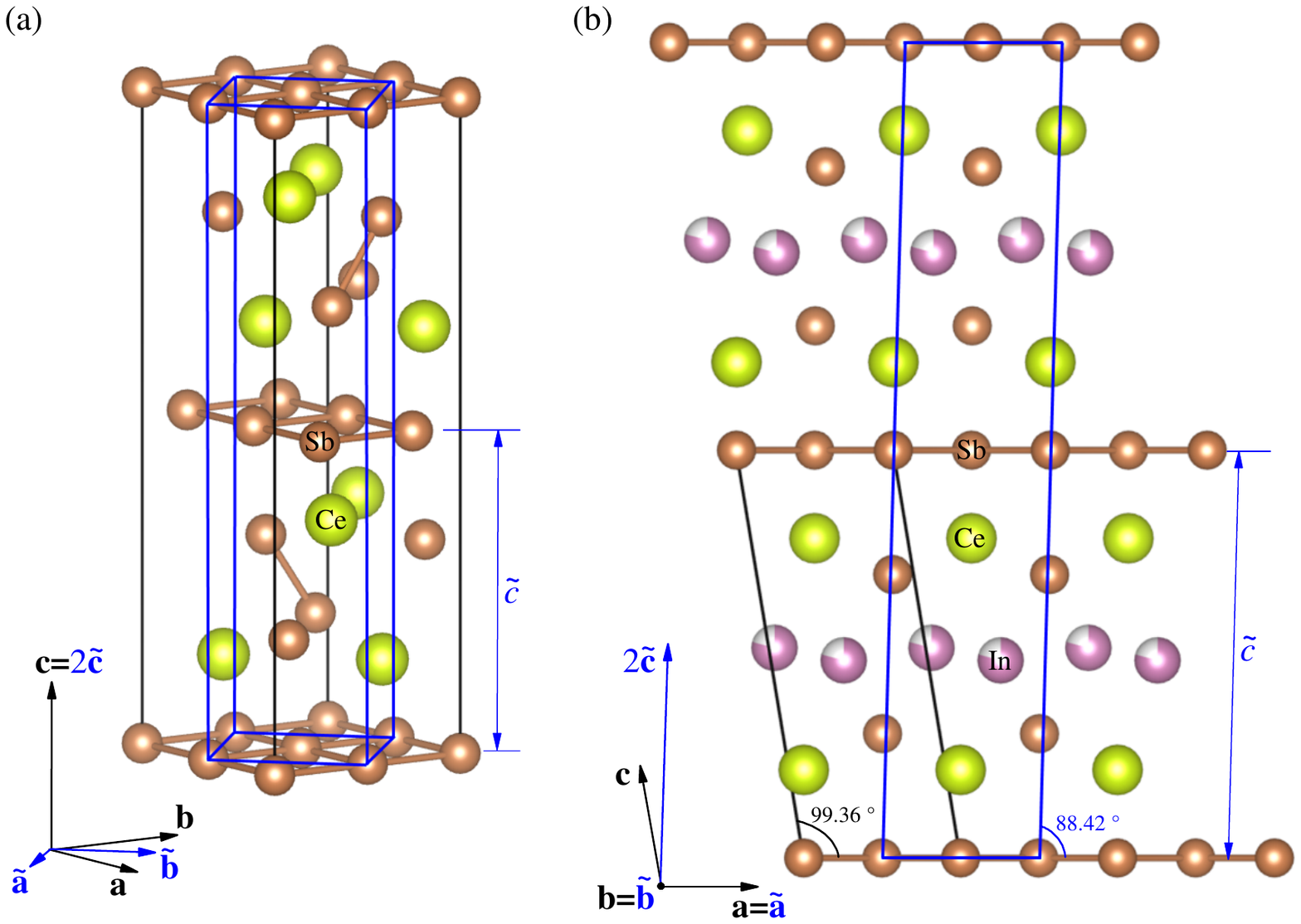}
\vspace*{-0pt}
\caption{\label{FigS1} Crystalline structure of CeSb$_2$ and CeIn$_{1-\delta}$Sb$_2$. To make a better comparison with other members of the Ce$Tm$Sb$_2$ family (Figure \ref{Fig1}), we modified the ``unit" cell. The original unit cell is depicted by the black lines, while the modified ``unit" cell is denoted by the blue lines. (a) CeSb$_2$ crystallizes in an orthorhombic structure with $a=6.295(6)$ \AA, $b=6.124(6)$ \AA, and $c=18.21(2)$ \AA~\cite{Wang-IC1967}. The modified $\tilde{\mathbf{a}}=(\mathbf{a}-\mathbf{b})/2$, $\tilde{\mathbf{b}}=(\mathbf{a}+\mathbf{b})/2$, $\tilde{\mathbf{c}}=\mathbf{c}/2$, and $\tilde{\gamma}=88.40^\circ$. (b) CeIn$_{1-\delta}$Sb$_2$ crystallizes a monoclinic structure with $a=4.478(2)$ \AA, $b=4.323(2)$ \AA,  $c=11.796(5)$ \AA, and $\beta=99.36~^\circ$ \cite{Ferguson-IC1999}. The modified ``unit" cell is quasi-orthorhombic, with $\tilde{\mathbf{a}}=\mathbf{a}$, $\tilde{\mathbf{b}}=\mathbf{b}$, and $\tilde{\beta}=88.42~^\circ$.}
\end{figure*}


\begin{table*}[!htp]
\caption{Crystalline structure parameters of CeSb$_2$ and Ce$Tm$Sb$_2$. $\tilde{c}$ characterizes the distance between adjacent Sb1 square nets, while $\tilde{a}$ and $\tilde{b}$ are the in-plane lattice parameters.}
\begin{ruledtabular}
\begin{tabular}{ccccccccccc}
Compound            & Space group & $c$ (\AA) & $\tilde{a}$ (\AA) $^\dag$ & $\tilde{b}$ (\AA) & $\tilde{c}$ (\AA) & $2\tilde{c}/(\tilde{a}+\tilde{b})$ \\ \hline
CeSb$_2$ \cite{Wang-IC1967}          & Cmca        &   18.21(2)         & 4.392(3)                  & 4.392(3)          & 9.11(1)          & 2.073                 \\
CeAgSb$_2$ \cite{Araki-PRB2003}        & P4/nmm      &  10.708(1)         & 4.3675(4)                 & 4.3675(4)         & 10.708(1)         & 2.452                 \\
CeIn$_{0.8}$Sb$_2$ \cite{Ferguson-IC1999} & P2$_1$/m    &  11.796(5)         & 4.478(2)                  & 4.323(2)          & 11.643(5)         & 2.646                 \\
CeSn$_{0.75}$Sb$_2$  & Cmcm        &   23.25(4)        & 4.239(9)                 & 4.52(3)        &  11.62(2)        & 2.654                 \\
\end{tabular}
\end{ruledtabular}
\small
\vspace*{-10pt}
\begin{flushleft}
\justifying{
$^{\dag}$ For CeSb$_2$, $a=6.295(6)$ \AA, $b=6.124(6)$ \AA, and $a~(b)\approx\sqrt{2}\tilde{a}~(\tilde{b})$. For the other three compounds, $a=\tilde{a}$, and $b=\tilde{b}$.
}
\end{flushleft}
\normalsize
\end{table*}

\newpage
\section{SM \Rmnum{2}. S\lowercase{ample characterization of \uppercase{C}e\uppercase{S}n$_{0.75}$\uppercase{S}b$_2$}}

\begin{figure*}[!ht]
\vspace*{-20pt}
\includegraphics[width=16cm]{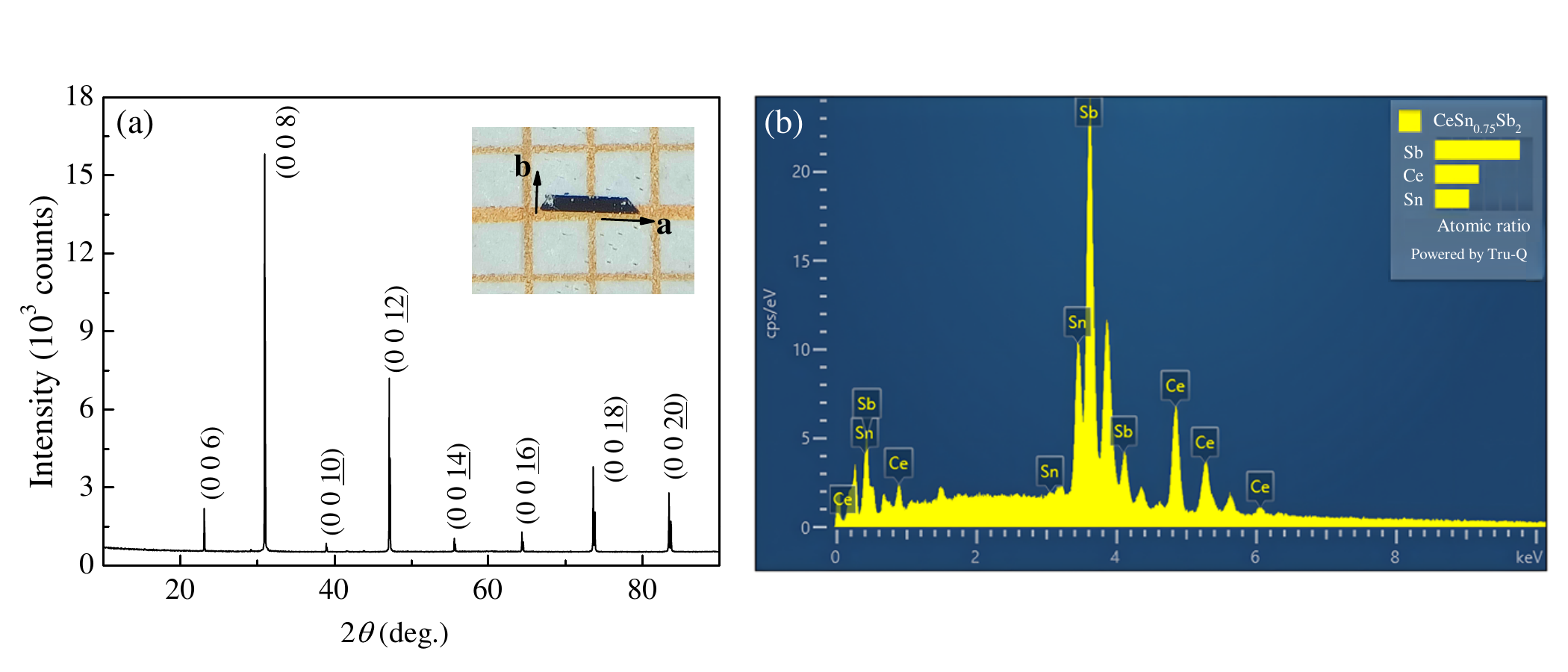}
\vspace*{-10pt}
\caption{\label{FigS2} (a) X-ray diffraction (XRD) pattern of the single-crystal CeSn$_{0.75}$Sb$_2$. Only $[0~0~2l]$ ($l=1,2...$) peaks are observed. The inset displays a photograph of the as-grown single crystal on millimeter-grid paper. (b) Energy-dispersive X-ray spectroscopy (EDS) measurements of CeSn$_{0.75}$Sb$_2$.}
\end{figure*}

\section{SM \Rmnum{3}. A\lowercase{dditional magnetization data}}

\begin{figure*}[!ht]
\vspace*{-0pt}
\includegraphics[width=16cm]{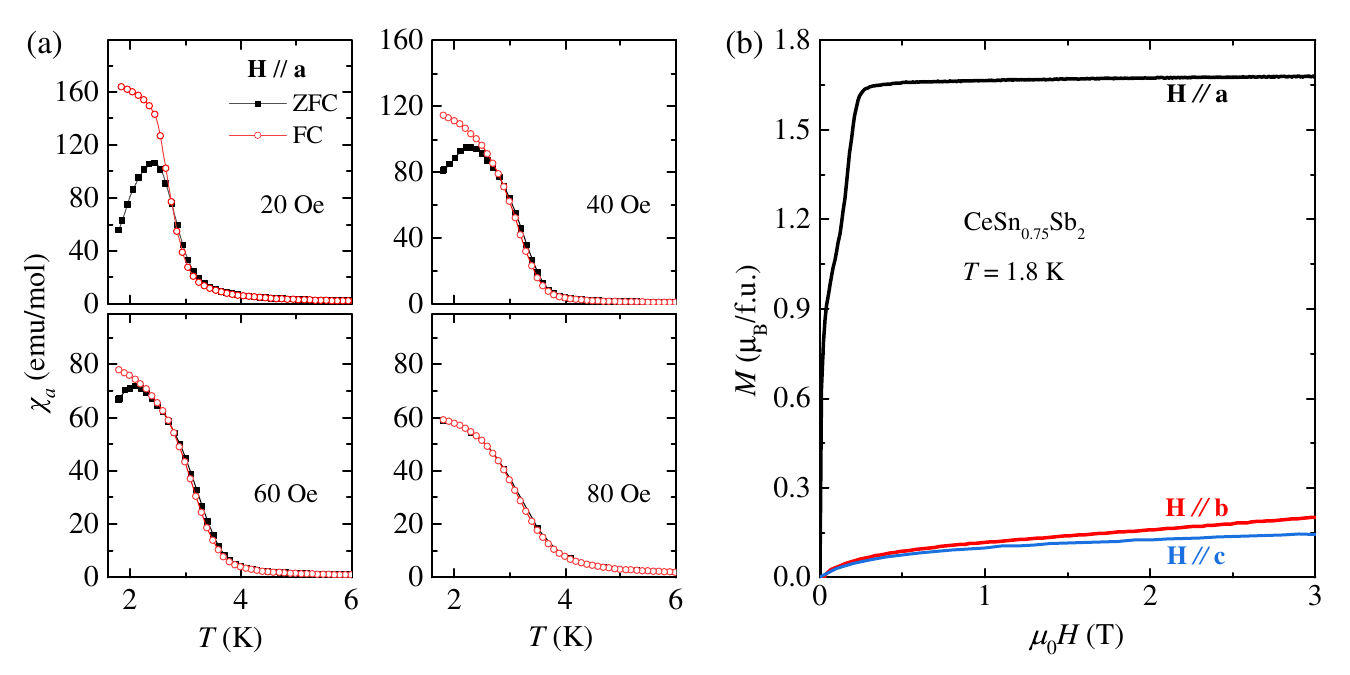}
\vspace*{-10pt}
\caption{ (a) The low-temperature $\chi (T)$ curves with $\mathbf{H}\parallel\mathbf{a}$ under tiny magnetic fields in zero-field-cooled (ZFC) and field-cooled (FC) modes. (b) Magnetization curves at 1.8 K for different crystallographic orientations. }
\label{FigS3}
\end{figure*}

The $\mathbf{a}$-axis magnetic susceptibility measured in zero-field-cooled (ZFC) and field-cooled (FC) protocols exhibit distinct irreversibility at low fields [Fig.~\ref{FigS3}(a)]. Increasing the external magnetic field suppresses this bifurcation. When the applied field reaches 80 Oe, the ZFC and FC curves overlap. This indicates the sensitivity of the cluster glass state to external magnetic fields.

Isothermal field dependent magnetization $M(H)$ further confirms $\mathbf{a}$ as the easy axis, as is shown in Fig.~\ref{FigS3}(b). For $\mathbf{H}\parallel\mathbf{a}$, $M(H)$ saturates at a small field $\sim$0.25 T after passing through a MMT transition. The saturated moment is 1.68 $\mu_B$/f.u. The out-of-plane magnetization is significantly smaller, measuring only 0.14 $\mu_\text{B}$/f.u at 3 T, consistent with a $|\pm 1/2\rangle$-dominant ground doublet. The $\mathbf{b}$-axis magnetization is slightly larger than $M_c$, but much smaller than $M_a$. This provides additional evidence for the abnormal magnetic anisotropy in CeSn$_{0.75}$Sb$_2$ that requires further investigations in the future.

\section{SM \Rmnum{4}. H\lowercase{all effect}}

\begin{figure}[!ht]
\vspace*{-10pt}
\includegraphics[width=14cm]{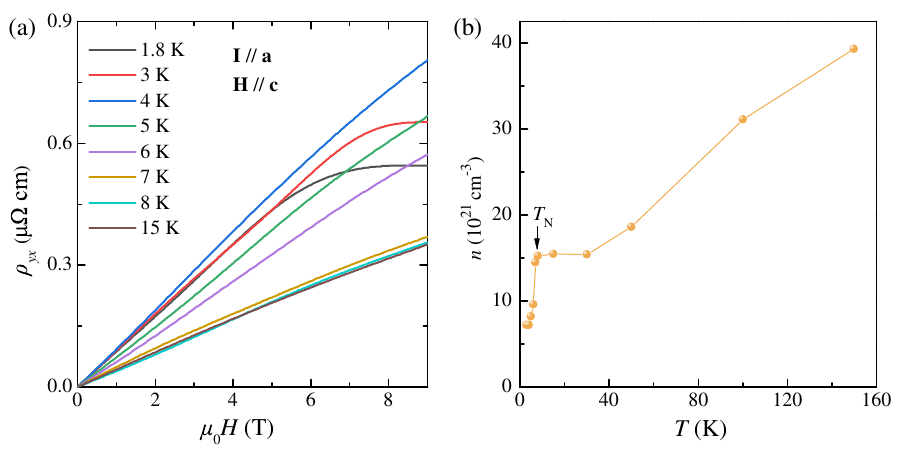}
\vspace*{-00pt}
\caption{Hall effect of CeSn$_{0.75}$Sb$_2$ for $\mathbf{H}\parallel\mathbf{c}$. (a) Field dependent Hall resistivity ($\rho_{yx}$) at selected temperatures. (b) Carrier density $n$ estimated from the Hall number in a single-band model. $n$ decreases sharply below $T_\text{N}$.}
\label{FigS4}
\end{figure}

\section{SM \Rmnum{5}. M\lowercase{agnetic entropy}}

\begin{figure}[!ht]
\vspace*{-20pt}
\includegraphics[width=9.5cm]{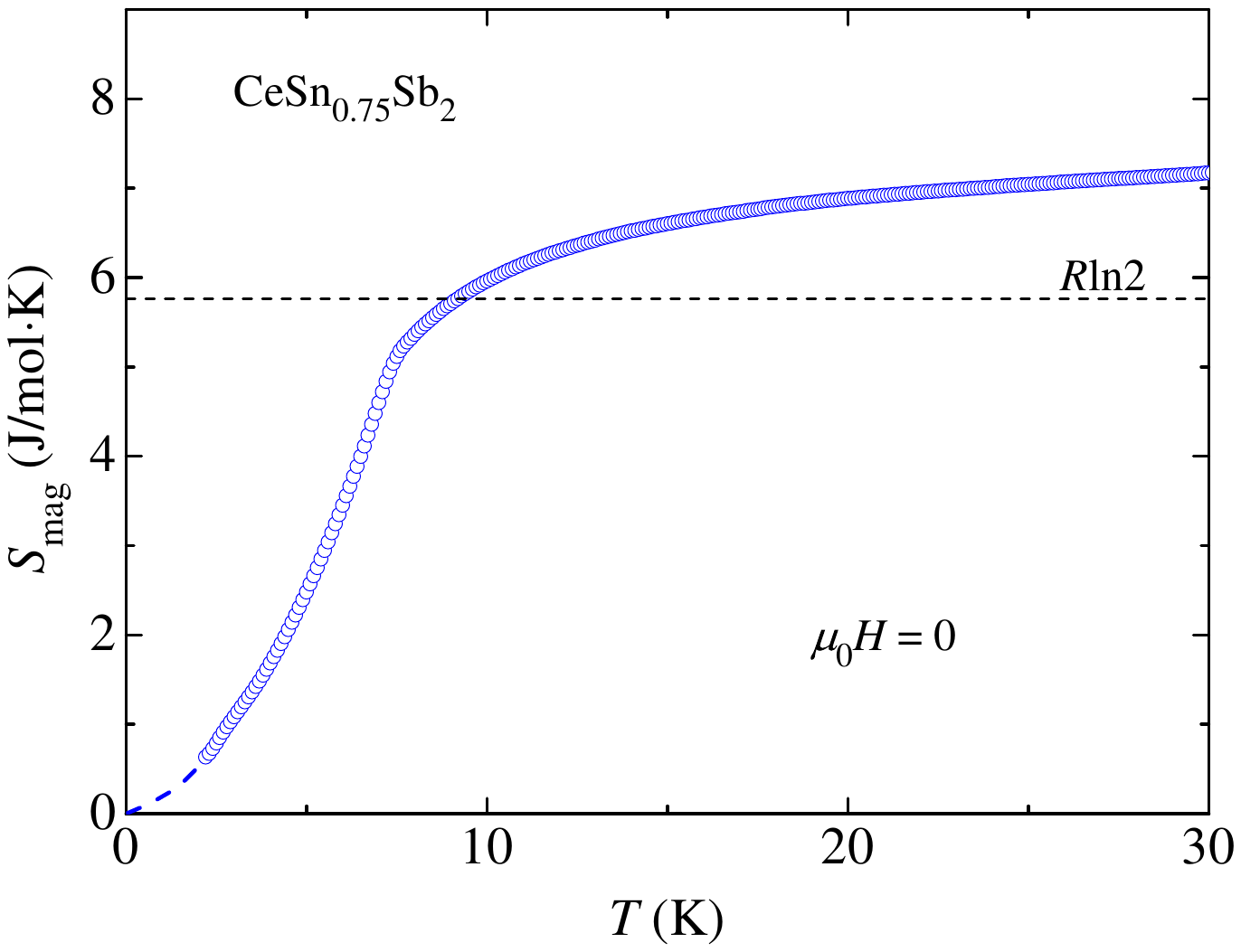}
\vspace*{-10pt}
\caption{Magnetic entropy gain of CeSn$_{0.75}$Sb$_2$ as a function of $T$. The dashed horizontal line depicts the value of $R\ln{2}$.}
\label{FigS5}
\end{figure}

\begin{figure}[!ht]
\vspace*{-00pt}
\includegraphics[width=16cm]{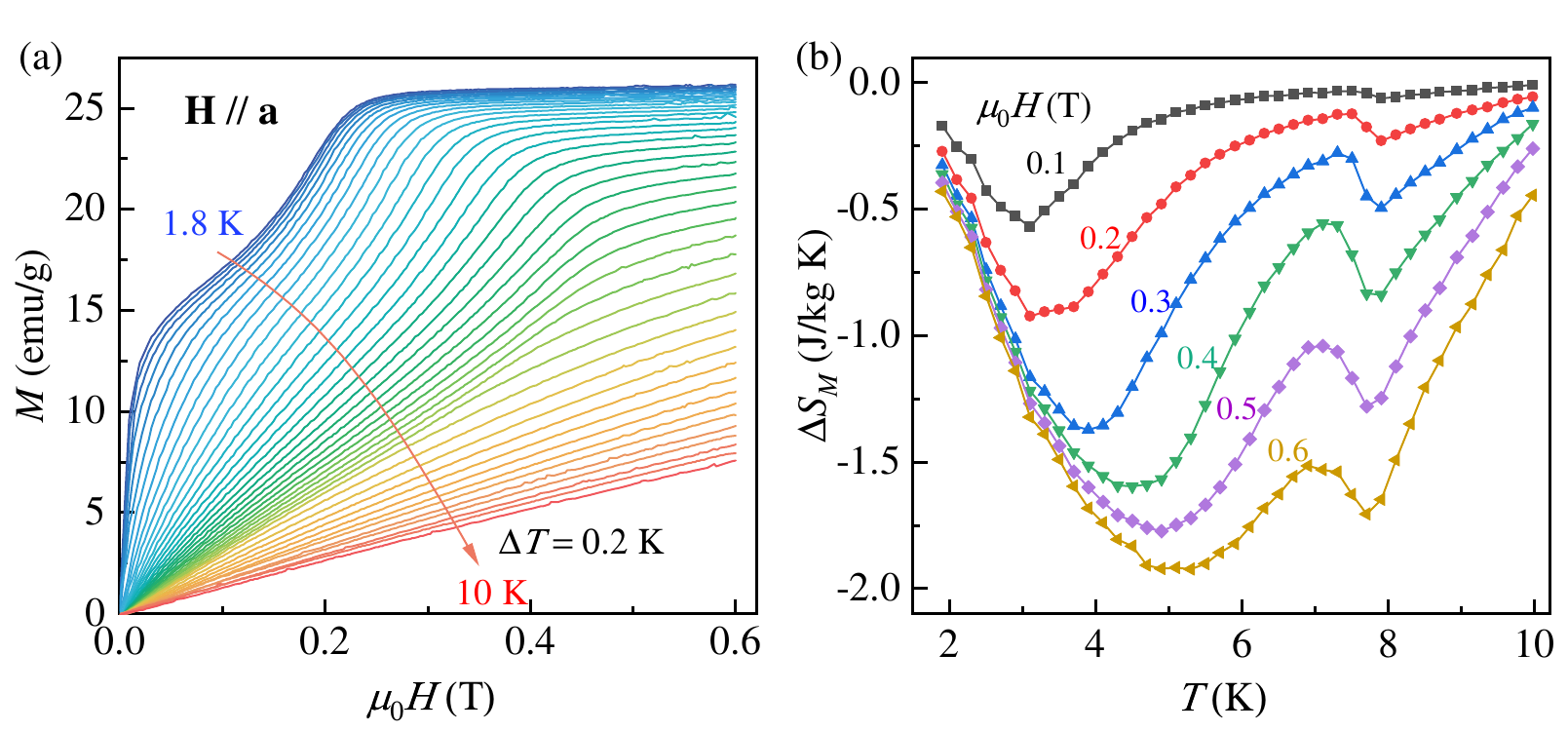}
\vspace*{-0pt}
\caption{(a) Isothermal magnetization curves of CeSn$_{0.75}$Sb$_2$ and (b) the calculated magnetic entropy change $\Delta S_\text{M}$ as a function of temperature. $\Delta S_\text{M}(H,T)=\int_0^H(\partial M/\partial T)|_H dH$. Taking $\mu_0 H=0.4$ T as an example, as the system cools into the AFM phase, the magnetic entropy exhibits an anomalous peak. Upon further cooling, it transits into the PPM state, where the entropy begins to decrease. This behavior corresponds to a counter-intuitive transition from an ordered, high-entropy state to a disordered, low-entropy state, which aligns with the thermodynamic criterion for inverse melting: the ordered phase possesses higher entropy than the disordered phase. Notably, for fields exceeding 0.2 T, the low-temperature entropy curves merge into a single trajectory, indicative of a field-polarized cluster-glass state.}
\label{FigS6}
\end{figure}

\end{document}